\documentclass[twoside,twocolumn,english,prc,amsmath,a4paper,myheadings]{revtex4}
\usepackage[T1]{fontenc}
\usepackage{geometry}
\usepackage{amsmath}
\usepackage{graphicx}
\usepackage{amssymb}

\makeatletter
\def\@oddhead{\rightmark \hfill  Ridges and Soft Jet Components in Untriggered Di-hadron Correlations  \hfill \thepage}
\def\@evenhead{\thepage \hfill K. Werner et al.\hfill}
\def\fnum@table{\tablename~{\bf\thetable}}
\def\fnum@figure{\figurename~{\bf\thefigure}}
\def\tablename{\footnotesize{\bf Table}}
\def\figurename{\footnotesize{\bf Figure}}

\usepackage{dcolumn}

\def\citet{\cite}

\makeatother

\usepackage{babel}

\begin{document}

\title{Ridges and Soft Jet Components in Untriggered Di-hadron Correlations
in Pb+Pb Collisions at 2.76 TeV}

\author{{\normalsize K.$\,$Werner$^{(a)}$, Iu.$\,$Karpenko$^{(b)}$, K.
Mikhailov$^{(c)}$, T.$\,$Pierog$^{(d)}$}}

\address{$^{(a)}$ SUBATECH, University of Nantes -- IN2P3/CNRS-- EMN, Nantes,
France}

\address{$^{(b)}$ Bogolyubov Institute for Theoretical Physics, Kiev 143,
03680, Ukraine}

\address{$^{(c)}$ Institute for Theoretical and Experimental Physics, Moscow,
117218, Russia}

\address{$^{(d)}$Karlsruhe Institute of Technology (KIT) - Campus North,
Institut f. Kernphysik, Germany}

\begin{abstract}
We study untriggered di-hadron correlations in Pb+Pb at 2.76 TeV,
based on an event-by-event simulation of a hydrodynamic expansion
starting from flux tube initial conditions. The correlation function
shows interesting structures as a function of the pseudorapidity difference
$\Delta\eta$ and the azimuthal angle difference $\Delta\phi$, in
particular comparing different centralities. We can clearly identify
a peak-like nearside structure associated with very low momentum components
of jets for peripheral collisions, which disappears towards central
collisions. On the other hand, a very broad ridge structure from asymmetric
flow seen at central collisions, gets smaller and finally disappears
towards peripheral collisions.
\end{abstract}
\maketitle
Two-dimensional di-hadron correlations provide a wealth of information
about the reaction dynamics of heavy ion collisions and proton-proton
scatterings. Experimental results have been obtained for Au+Au collisions
at 200 GeV\citet{ridge1,ridge2,ridge3} and for proton-proton reactions
at 7 TeV \citet{ridge4}, results for Pb+Pb will appear soon. Whereas
in most applications momentum triggers are employed, we will discuss
in this letter untriggered correlations, dominated by very low momentum
pairs. Also here, one observes a nearside ridge-like structure extended
over many units in $\Delta\eta$. In this letter, we will discuss
the centrality dependence of the two-dimensional di-hadron correlation
function in Pb+Pb collisions at 2.76 TeV. 

\begin{figure}[tb]
\begin{centering}
\includegraphics[scale=0.3]{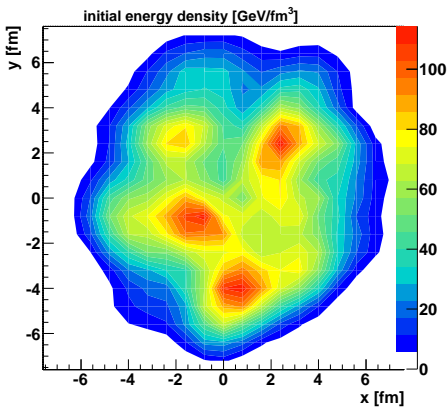}
\par\end{centering}

\caption{Initial energy density as a function of the transverse coordinates,
at space-time rapidity $\eta_{s}=0$.\label{fig:eps}}

\end{figure}
\begin{figure}[b]
\begin{centering}
\includegraphics[scale=0.3]{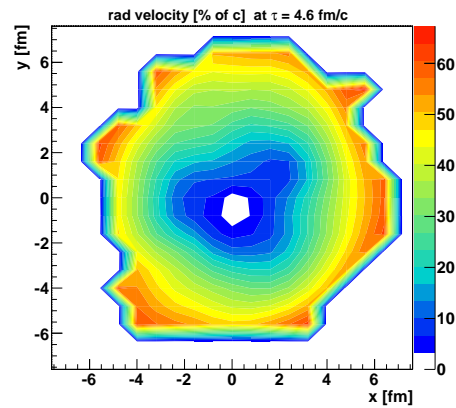}
\par\end{centering}

\caption{Radial flow velocity (in \% of $c$) as a function of the transverse
coordinates, at space-time rapidity $\eta_{s}=0$, at $\tau=4.6\,$fm/c.\label{fig:rad}}

\end{figure}

We employ a sophisticated hydrodynamical scenario (for details see
\citet{epos2}), with initial conditions obtained from a flux tube
approach (EPOS), compatible with the string model, used since many
years for elementary collisions (electron-positron, proton proton),
and the color glass condensate picture \citet{cgc1}. The equation-of-state
is compatible with lattice gauge results of ref. \citet{lattice}.
We use a hadronic cascade procedure after hadronization from the thermal
system at an early stage \citet{urqmd,urqmd2}.

For the present discussion it is important to note that we perform
event-by-event simulations, taking into account the highly irregular
space structure of single events, as shown in fig. \ref{fig:eps}.
There are a couple of {}``hot spots'' visible, which have actually
a long range structure in longitudinal direction (a very similar picture
is obtained for different values of $\eta_{s}$ (longitudinal translational
invariance)). The irregular structure of the initial energy density
translates in an irregular transverse flow some time later, as seen
in fig. \ref{fig:rad}. One can easily see that the two remarkable
peaks in the lower half of the transverse plane in fig. \ref{fig:eps}
squeeze matter outwards with large velocity just in between them,
see fig. \ref{fig:rad}. 

\begin{figure*}[t]
\begin{centering}
\includegraphics[scale=0.38]{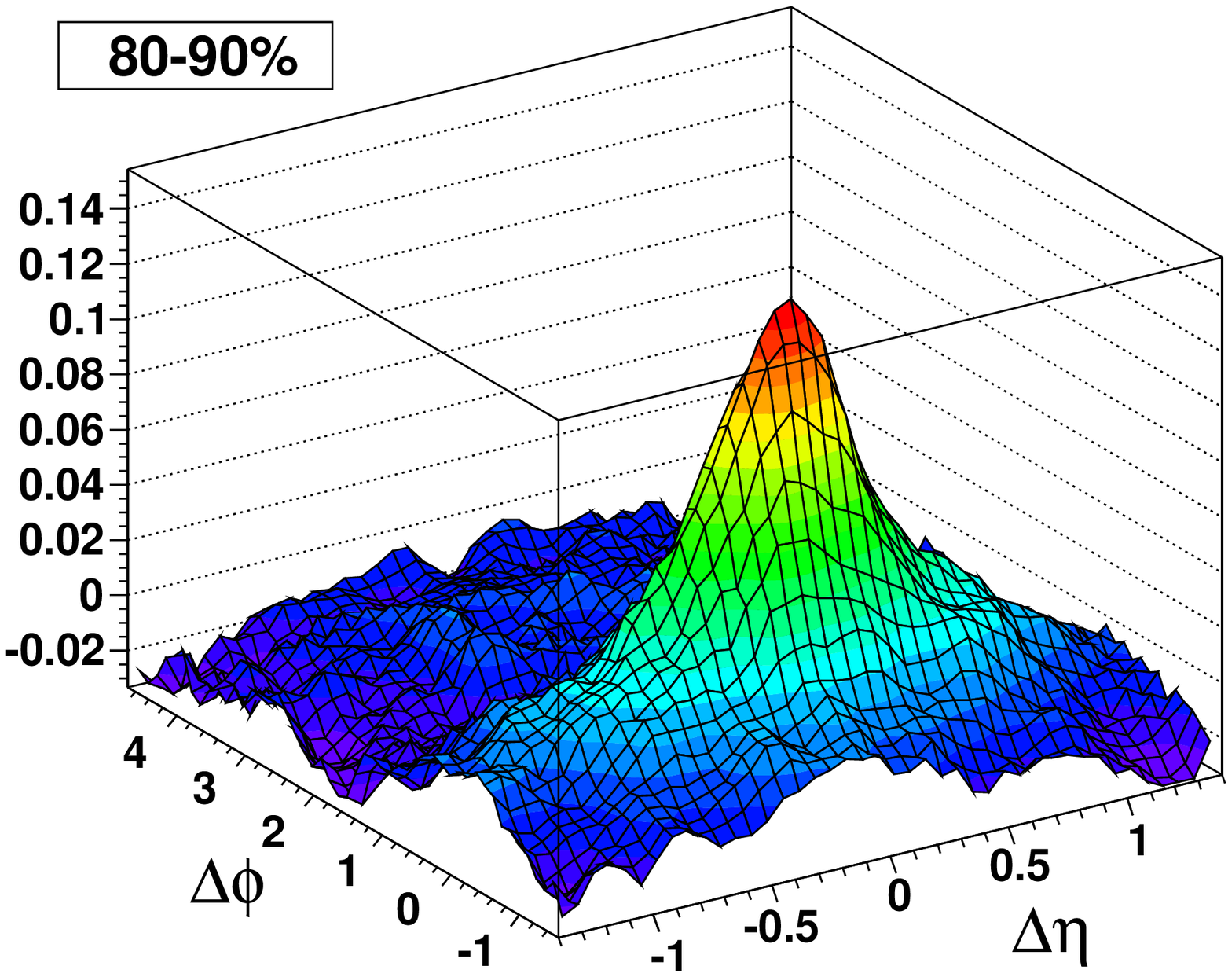}\includegraphics[scale=0.38]{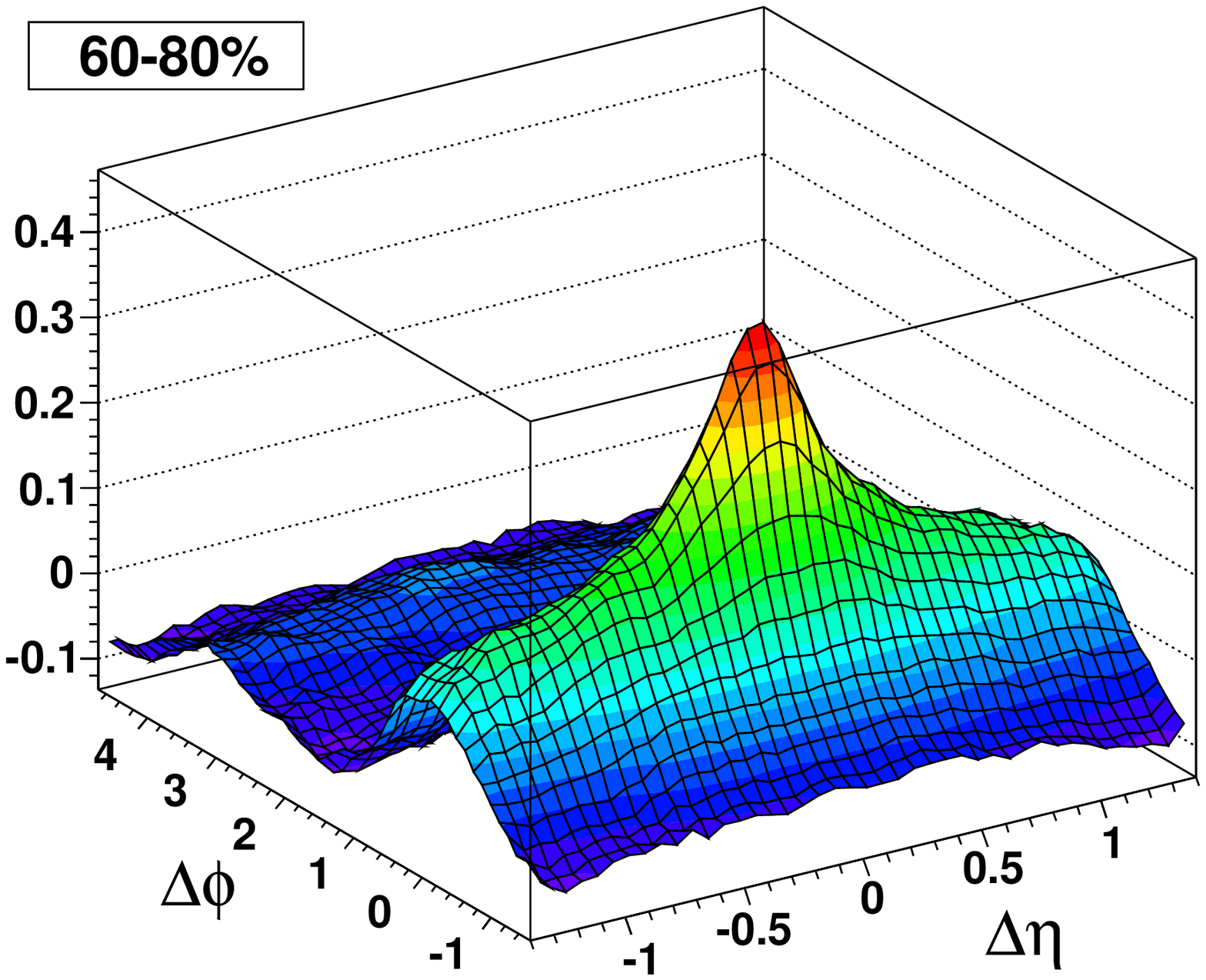}
\par\end{centering}

\begin{centering}
\vspace{-0.35cm}

\par\end{centering}

\begin{centering}
\includegraphics[clip,scale=0.38]{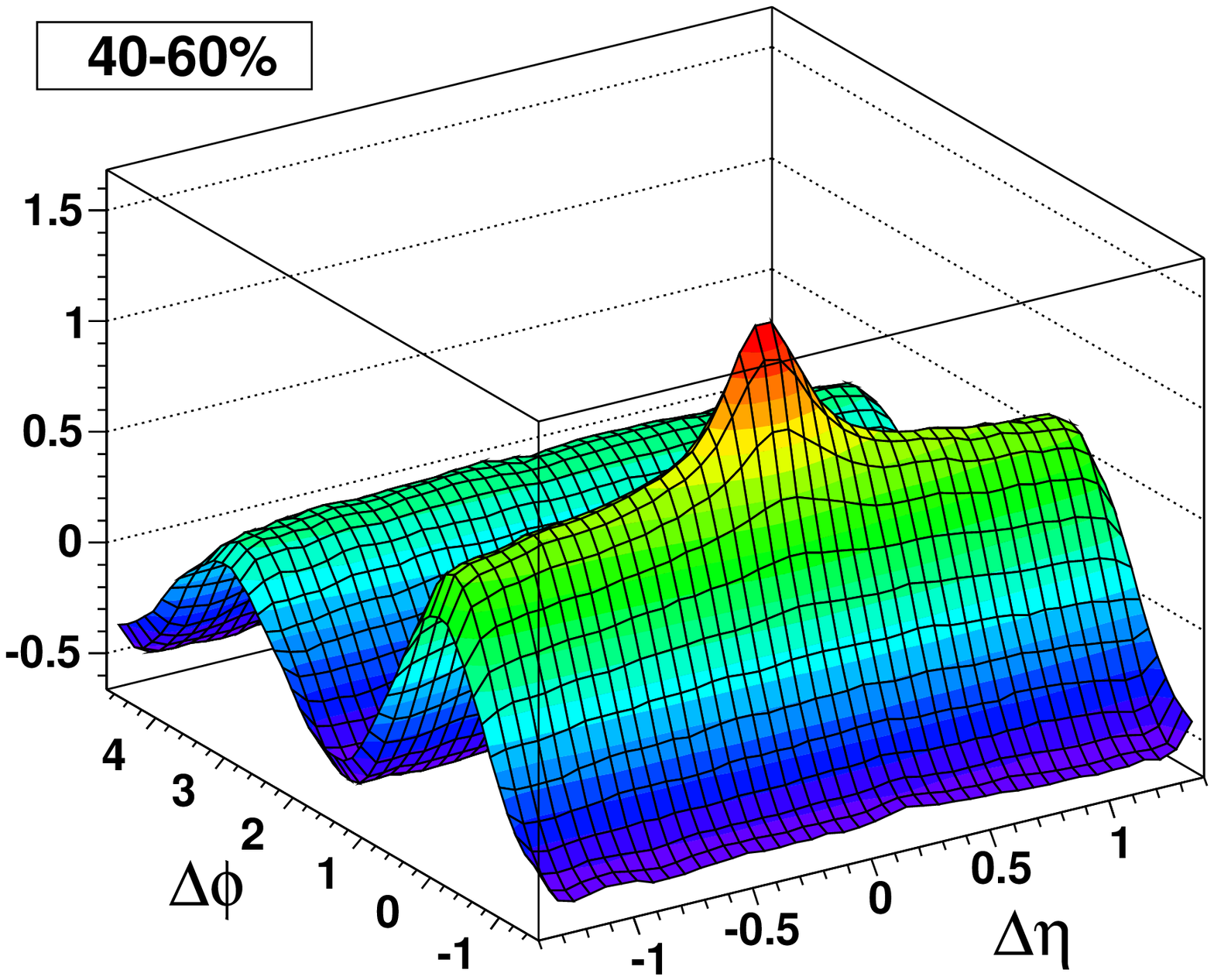}\includegraphics[clip,scale=0.38]{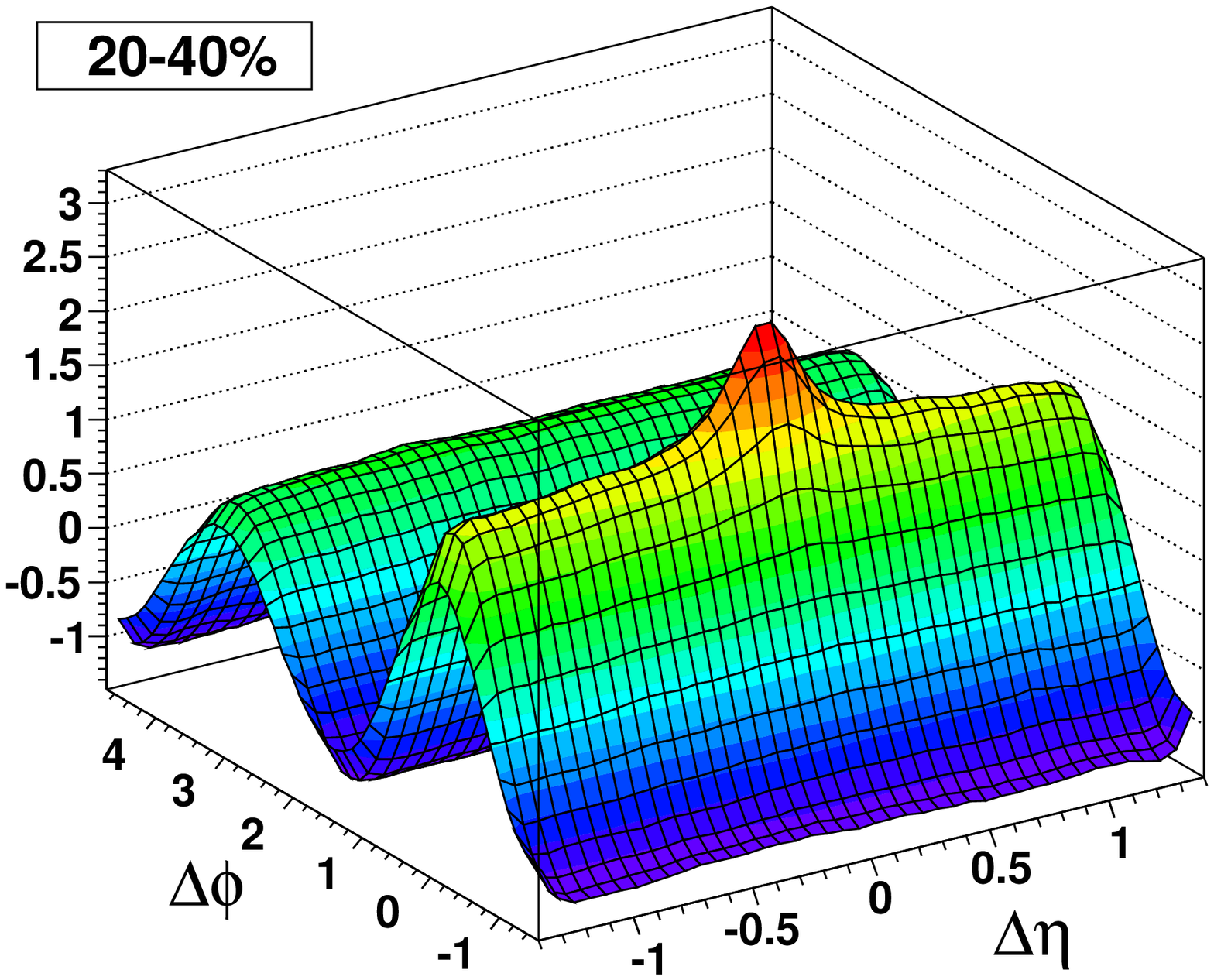}
\par\end{centering}

\vspace{-0.35cm}

\begin{centering}
\includegraphics[clip,scale=0.38]{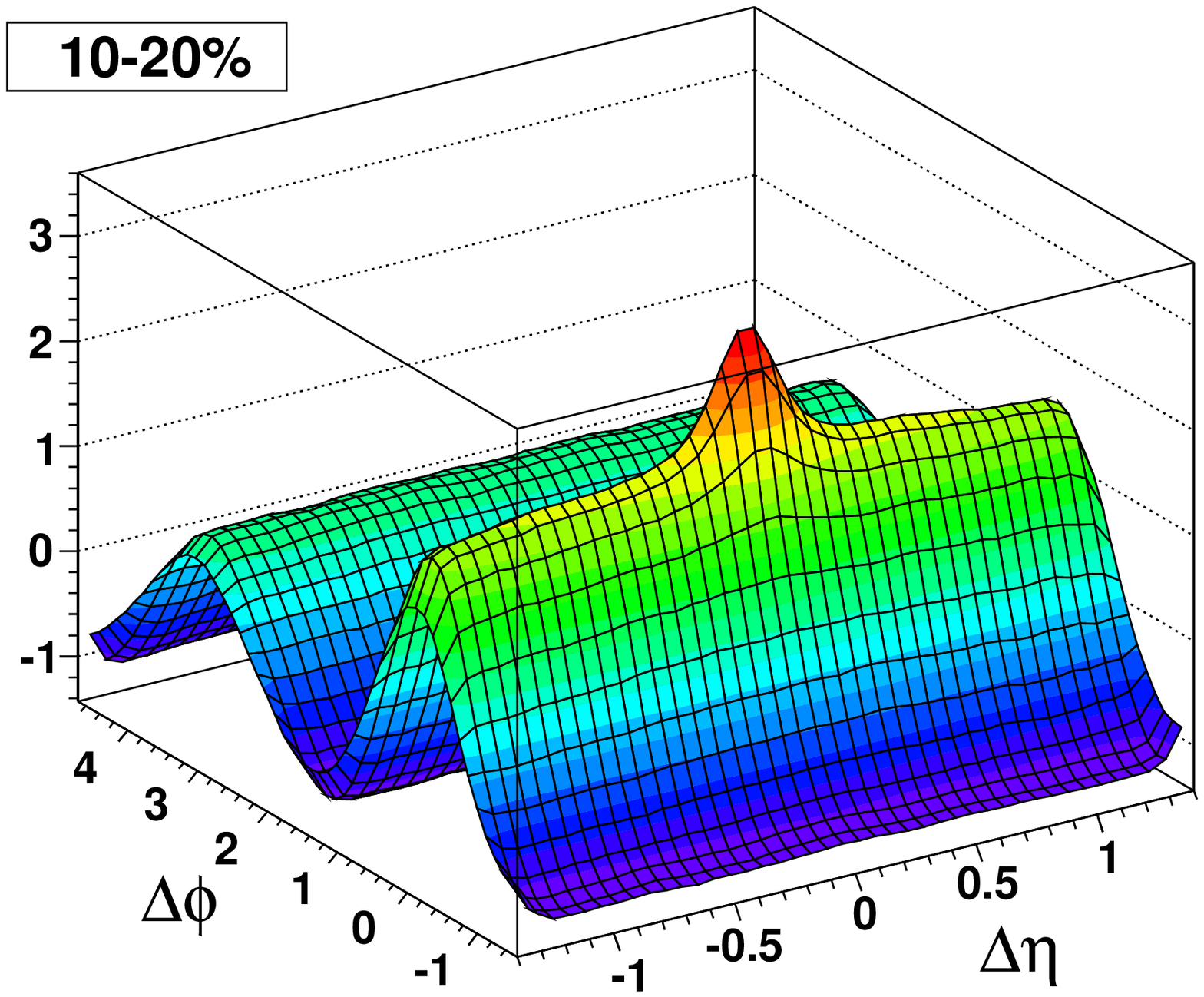}\includegraphics[clip,scale=0.38]{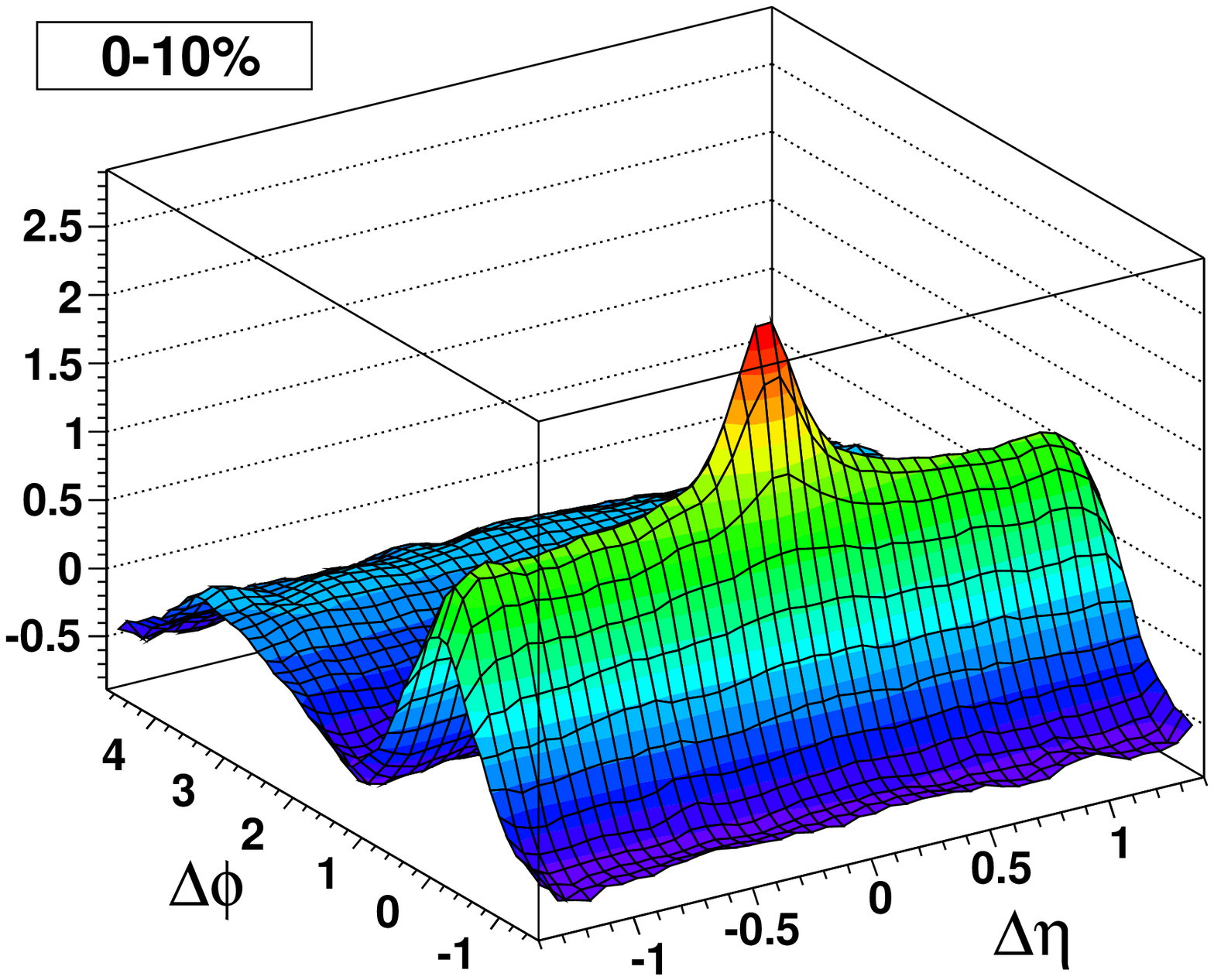}
\par\end{centering}

\caption{(Color online) Untriggered di-hadron correlation function $R(\Delta\eta,\Delta\phi)$
including Bose-Einstein statistics versus $\Delta\eta$ and $\Delta\phi$
for different centralities. \label{cap:ridge1} }

\end{figure*}
\begin{figure*}[t]
\begin{centering}
\includegraphics[scale=0.38]{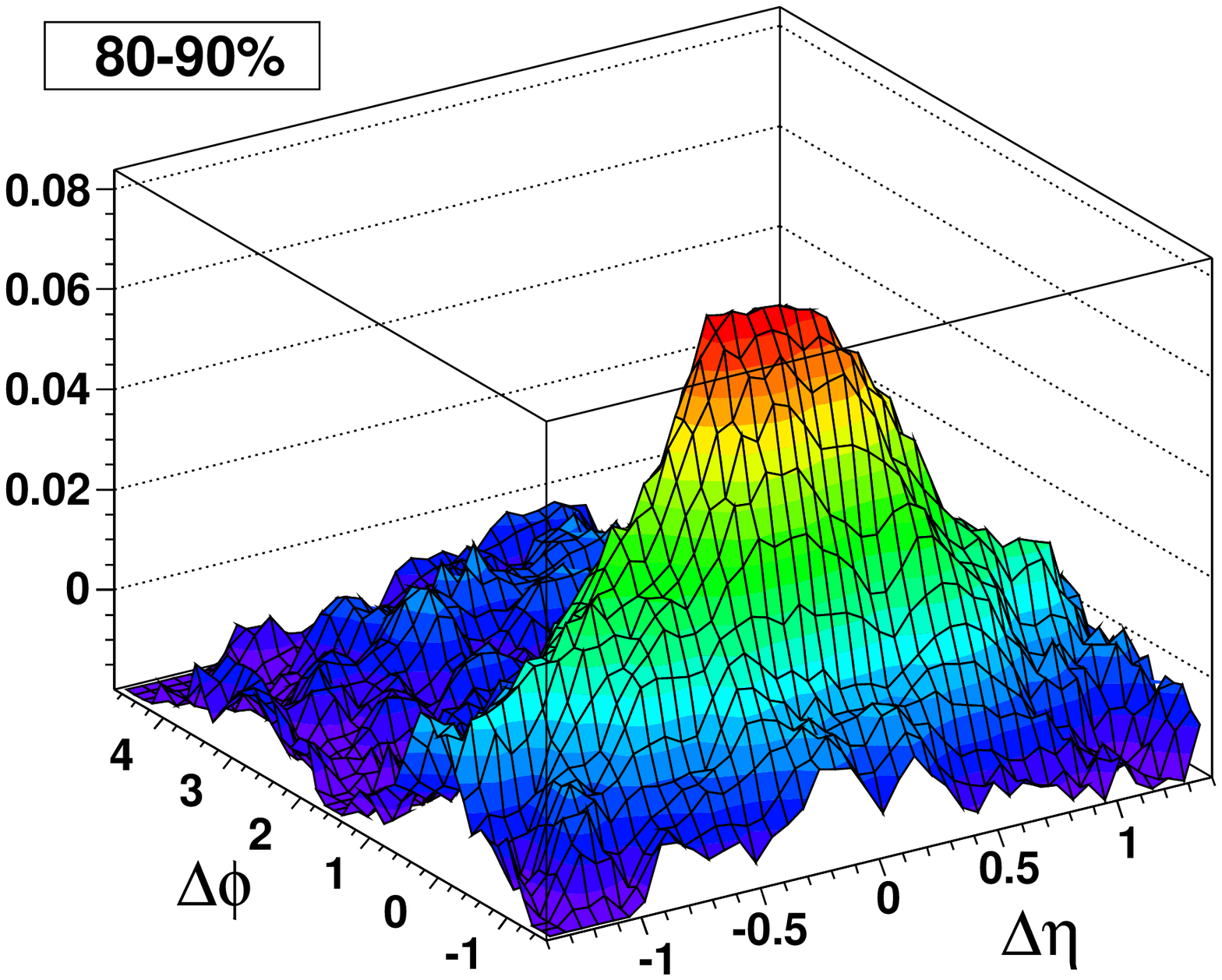}\includegraphics[scale=0.38]{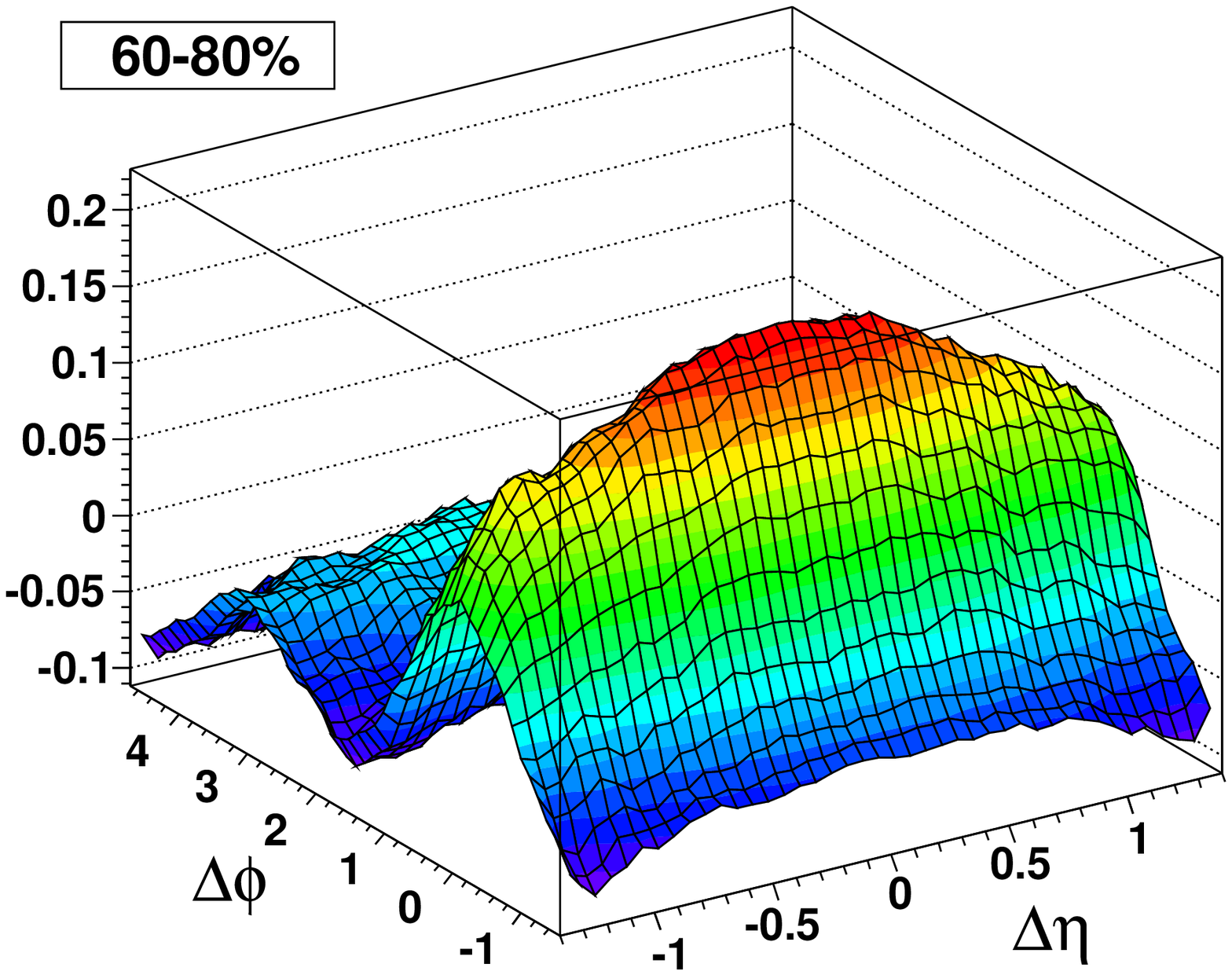}
\par\end{centering}

\begin{centering}
\vspace{-0.35cm}

\par\end{centering}

\begin{centering}
\includegraphics[clip,scale=0.38]{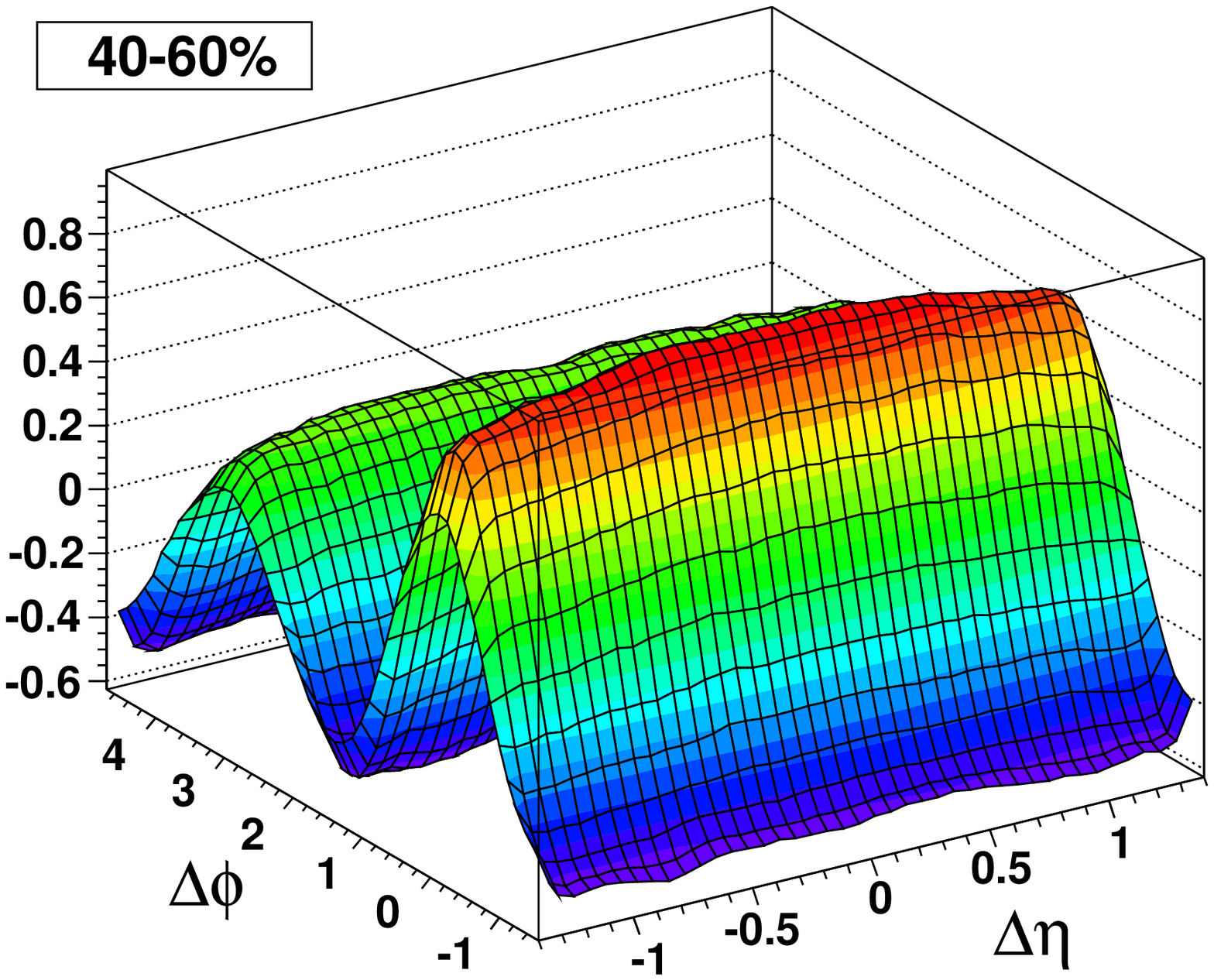}\includegraphics[clip,scale=0.38]{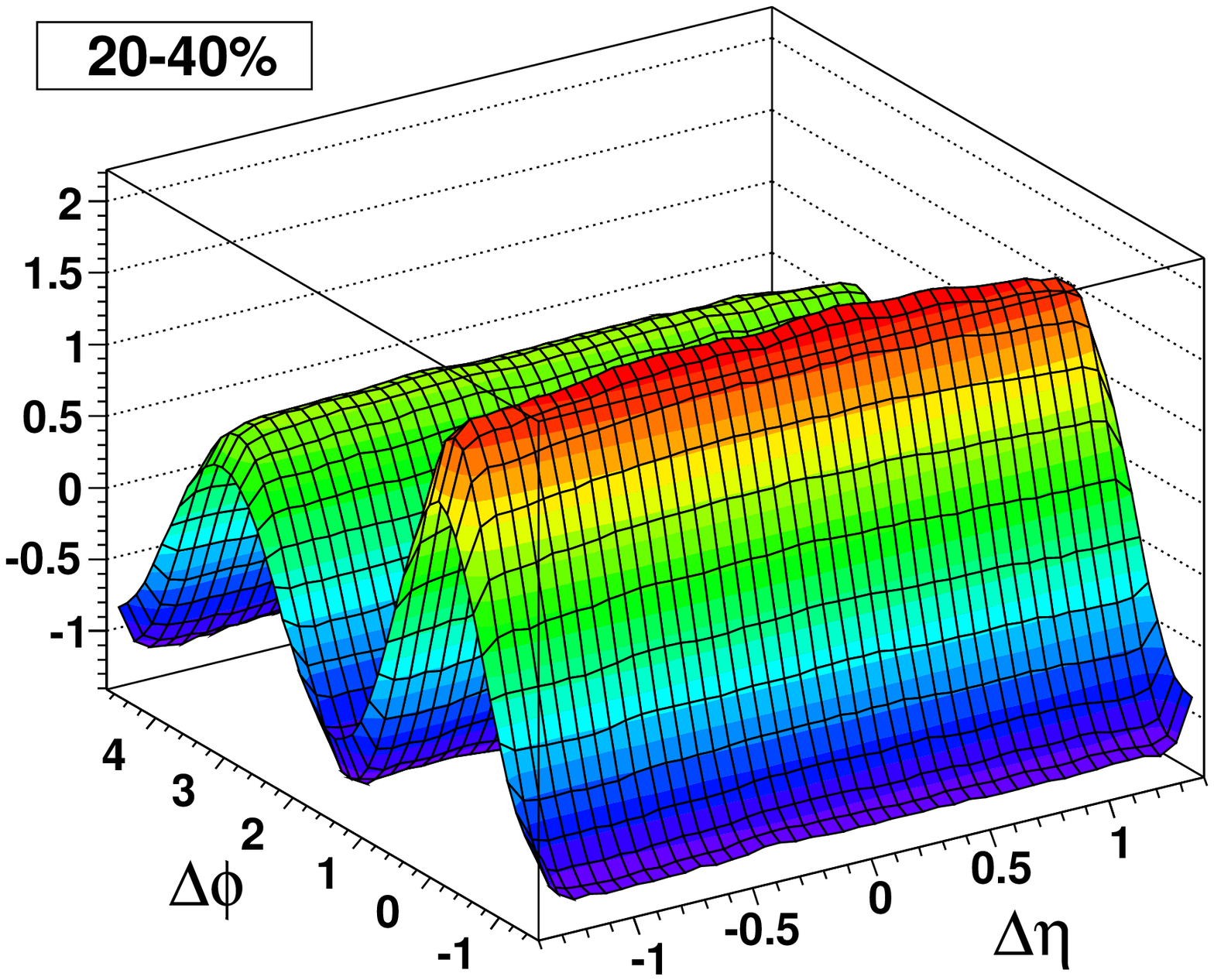}
\par\end{centering}

\vspace{-0.35cm}

\begin{centering}
\includegraphics[clip,scale=0.38]{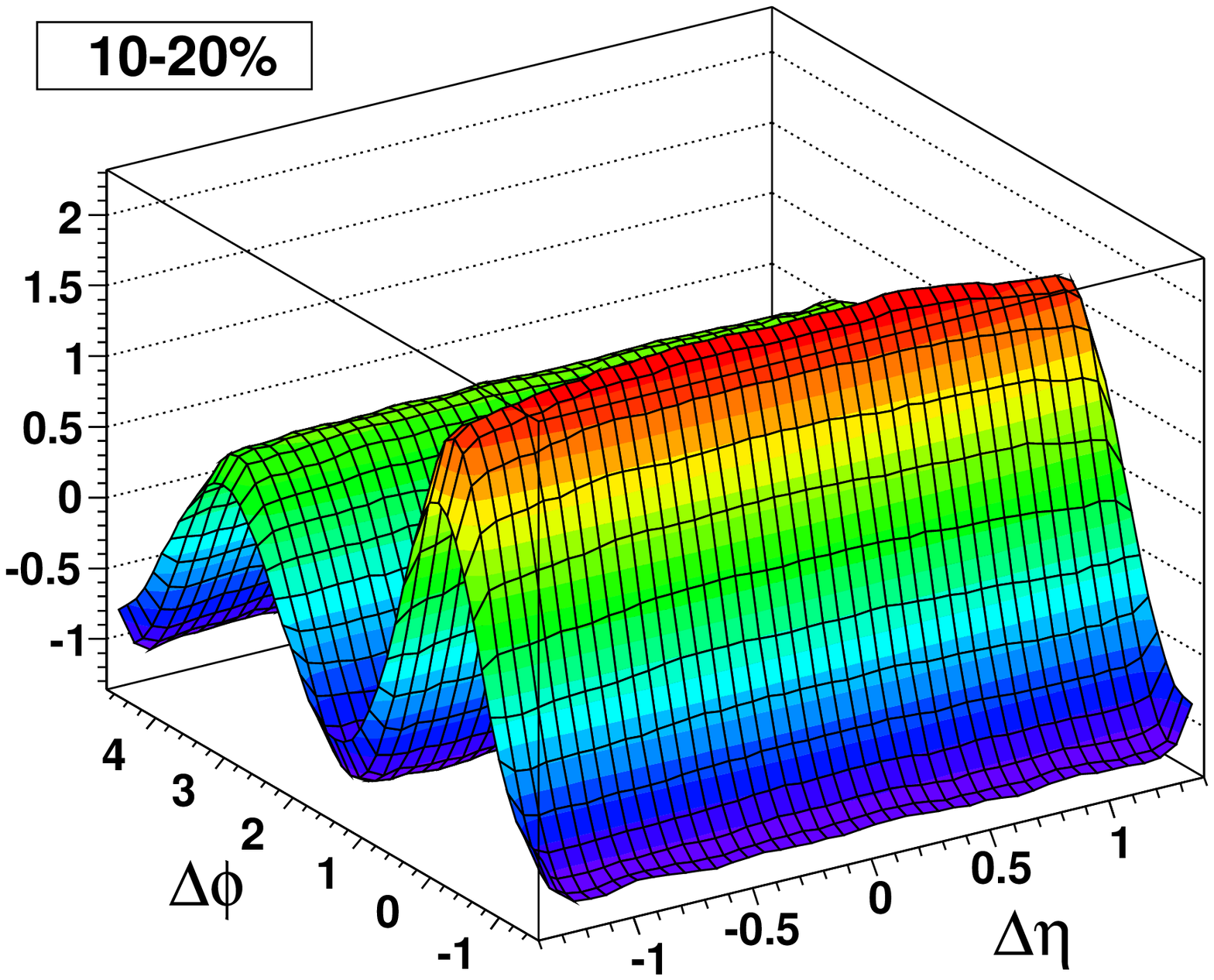}\includegraphics[clip,scale=0.38]{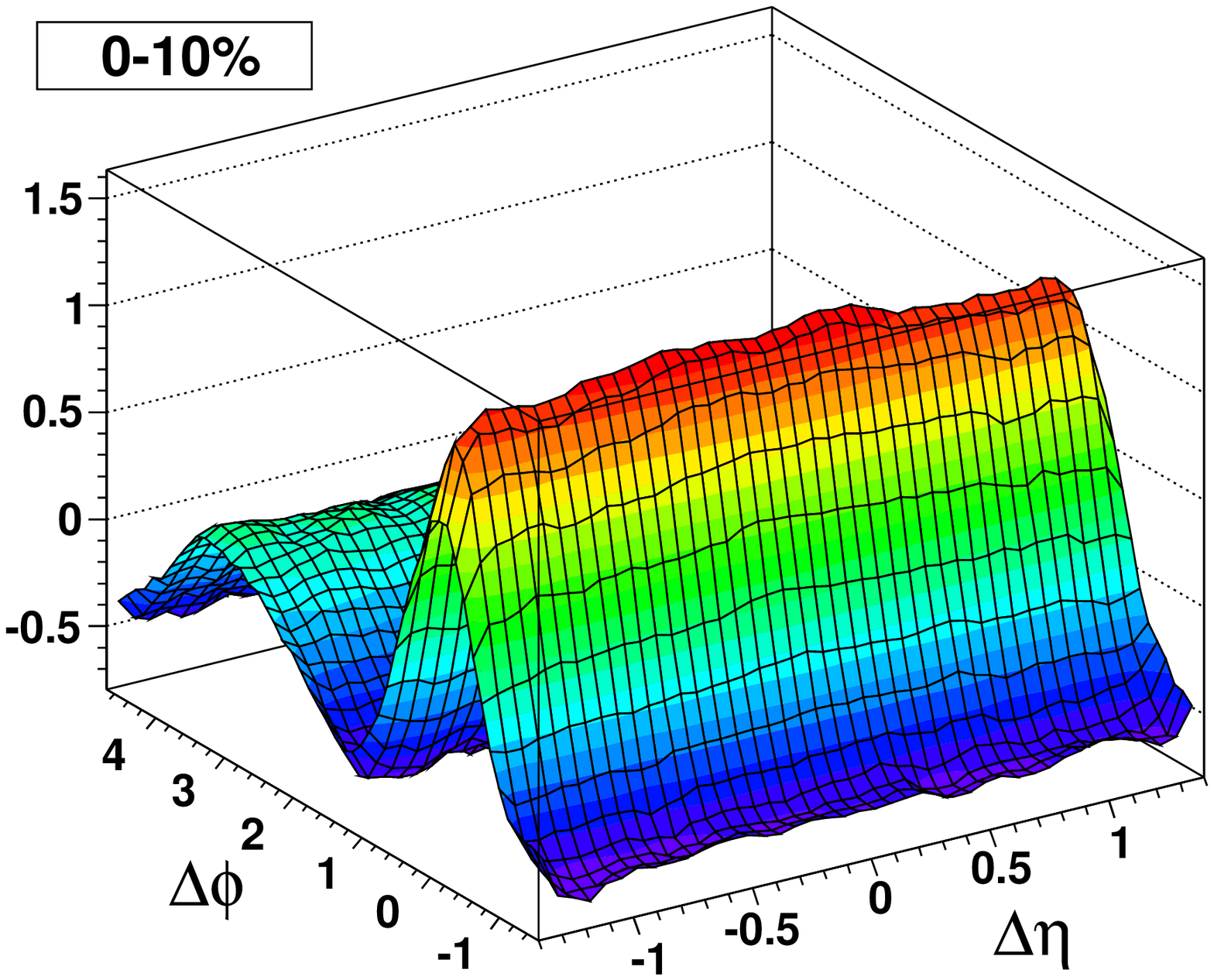}
\par\end{centering}

\caption{(Color online) Untriggered di-hadron correlation function $R(\Delta\eta,\Delta\phi)$
without Bose-Einstein statistics versus $\Delta\eta$ and $\Delta\phi$
for different centralities. \label{cap:ridge2} }

\end{figure*}
\begin{figure}[tb]
\begin{centering}
\includegraphics[angle=270,scale=0.25]{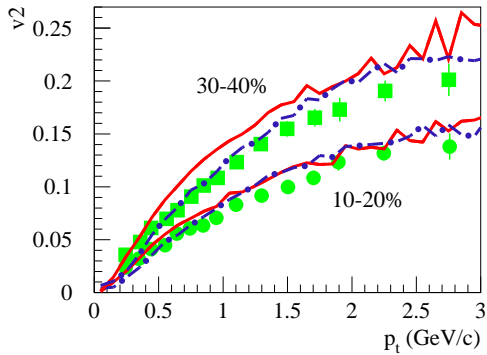}
\par\end{centering}

\caption{(Color online) The transverse momentum dependence of $v_{2}$ for
charged particles, compared to data (points), for two different centralities.
We show the full calculation (solid line), and a calculation without
hadronic cascade (dashed-dotted).\label{fig:v2}}

\end{figure}

Based on the the above scenario, we compute the two-dimensional di-hadron
correlations function $R$ as a function of the pseudorapidity difference
$\Delta\eta$ and the azimuthal angle difference $\Delta\phi$. We
use $R=$ $C\,(\rho_{\mathrm{real}}/\rho_{\mathrm{mixed}}-1)$, with
a normalization $C=N/(2\pi\Delta)$, where $N$ is the multiplicity
and $\Delta$the pseudorapidity range. We show in fig. \ref{cap:ridge1}
the results for different centralities, using a full calculation,
including Bose-Einstein statistics (for $\pi^{+}\pi^{+}$ and $\pi^{-}\pi^{-}$pairs),
and in fig. \ref{cap:ridge2} the corresponding results without Bose-Einstein
statistics.%
After removing the Bose-Einstein peak (in our case making the calculation
without Bose-Einstein statistics, see fig. \ref{cap:ridge2}) there
are three structures visible:

\begin{itemize}
\item The elliptical flow of the form $\cos(2\Delta\phi),$ strongest at
intermediate centralities, but also present for central collisions.
\item A very broad ridge at $\Delta\phi=0$, which gets smaller towards
more peripheral collisions, and disappears for the most peripheral
bin, showing a weak $\eta$ dependence.
\item A peak around $\Delta\phi=0,\:\Delta\eta=0$, very pronounced for
most peripheral collisions, getting weaker towards more central events,
and disappearing for central collisions.
\end{itemize}
The $\cos(2\Delta\phi)$ component seems to be independent of $\eta$,
so we can crosscheck the model by comparing with $v_{2}=\left\langle \cos2\phi\right\rangle $
measurements for $\eta=0$ \citet{ali3-flow} at two different centralities:
In fig. \ref{fig:v2}, we show the transverse momentum dependence
of $v_{2}$.

The ridge contribution is most easily discussed for the highest centrality
class (see fig. \ref{cap:ridge2}, 5-10\%), where {}``the ridge''
is the difference $R(\Delta\eta,\Delta\phi)-R(\Delta\eta,\pi)\cos2\Delta\phi$.
The reason for this correlation in our approach can be seen from figs.
\ref{fig:eps} and \ref{fig:rad}: individual events show typically
(due to random fluctuations) a certain number of {}``hot spots''
of very high energy density, elongated in longitudinal direction in
the string model approach ($\to$ {}``hot tubes''). This leads to
a squeeze-out of matter with high radial velocity at certain azimuthal
angles $\phi_{K}$, typically between the hot tubes. The effect get
weaker towards more peripheral collisions, because the transverse
area gets smaller, and there is finally only a single hot tube in
the center, which will not produce any ridge structure.

The peak in the peripheral centrality class (see fig. \ref{cap:ridge2},
80-90\%) is in our model clearly identified as coming from jets. It
should be said that at 2.76 TeV, in our model, all elementary interactions
are hard (so sloppy spoken: everything comes from {}``jets''). The
correspond elementary flux tubes are kinky strings, which are mainly
longitudinally, but there are transversely moving parts, carrying
the momenta of the hard scatterings (see ref. \citet{epos2}). These
strings are the basis for the calculation of the initial energy density,
and of course very high momentum string segments have to excluded.
We employ a somewhat modified procedure for bulk / jet separation
compared to our earlier work, where all string segment with $p_{t}>p_{t}^{\mathrm{cut}}$
could escape unmodified. Here we have in mind a picture where the
high transverse momentum string segments lose energy via the energy
loss of the corresponding partons, and therefore the energy loss of
the string segments moving a distance $dL$ through space characterized
by an energy density $\varepsilon$ (from other strings) is given
via the parton energy loss formula $\Delta E\propto\varepsilon^{3/8}\sqrt{E}dL$
\citet{baier}. For low transverse momentum segments, we use simply
$\Delta E\propto\rho dL$, with $\rho$ being the string density.
In any case, segments with $E>\Delta E$ escape the plasma -- which
is also possible for low momentum segments, sitting on the surface
of the matter distribution.

In this sense the peak in peripheral collisions is due to escaping
jet elements. The peak disappears for more central collisions, because
of an increase of the chance of hadronic rescattering of these low
momentum jet elements with frozen out particles from the plasma. 

In summary, untriggered two-dimensional di-hadron correlations provide
interesting information about a multitude of dynamical features of
the expanding plasma: the elliptical flow as a consequence of a global
azimuthal asymmetry of the initial matter distribution; the ridge
coming from initial density fluctuations, which lead to asymmetric
radial squeeze out of matter; a peak from low momentum jet components,
having survived the plasma -- providing a link to the crucial question
of separating bulk and the low momentum pieces of jets.

\end{document}